\documentclass[usenatbib,usegraphicx,fleqn]{mn2e}
\usepackage{aas_macros}
\usepackage{amsmath}
\usepackage{amssymb}
\usepackage{float}
\usepackage{times}
\usepackage[T1]{fontenc}
\usepackage{aecompl}

\begin{document}

\title[The soft state of Swift J1753.5-0127]{A low-luminosity soft state in the short period black hole X-ray binary Swift J1753.5-0127}
\author[A.W. Shaw et al.]{A.W. Shaw,$^{1}$\thanks{Email: A.Shaw@soton.ac.uk} P. Gandhi,$^1$ D. Altamirano,$^1$ P. Uttley,$^2$ J.A. Tomsick,$^3$ P.A. Charles,$^1$ F. F\"{u}rst,$^4$\newauthor{F. Rahoui,$^{5,6}$ and D.J. Walton$^{7,8}$} \\
$^{1}$Department of Physics and Astronomy, University of Southampton, Southampton SO17 1BJ, UK \\
$^{2}$Anton Pannekoek Institute, University of Amsterdam, Science Park 904, NL-1098 XH Amsterdam, the Netherlands \\
$^{3}$Space Sciences Laboratory, 7 Gauss Way, University of California, Berkeley, CA 94720-7450, USA \\
$^{4}$Cahill Center for Astronomy and Astrophysics, California Institute of Technology, Pasadena, CA 91125, USA \\
$^{5}$European Southern Observatory, K. Schwarzschild-Str. 2, 85748 Garching bei M\"unchen, Germany \\
$^{6}$Department of Astronomy, Harvard University, 60 Garden street, Cambridge, MA 02138, USA\\
$^{7}$Jet Propulsion Laboratory, California Institute of Technology, Pasadena, CA 91109, USA\\
$^{8}$Space Radiation Laboratory, California Institute of Technology, Pasadena, CA 91125, USA}

\date{\today}
\maketitle
\begin{abstract}
We present results from the spectral fitting of the candidate black hole X-ray binary Swift J1753.5-0127 in an accretion state previously unseen in this source. We fit the 0.7--78 keV spectrum with a number of models, however the preferred model is one of a multi-temperature disk with an inner disk temperature $\mathrm{k}T_\mathrm{in}=0.252\pm0.003$ keV scattered into a steep power-law with photon index $\Gamma=6.39^{+0.08}_{-0.02}$ and an additional hard power law tail ($\Gamma=1.79\pm0.02$). We report on the emergence of a strong disk-dominated component in the X-ray spectrum and we conclude that the source has entered the soft state for the first time in its $\sim$10 year prolonged outburst. Using reasonable estimates for the distance to the source ($3$ kpc) and black hole mass ($5M_{\odot}$), we find the unabsorbed luminosity (0.1--100 keV) to be $\approx0.60\%$ of the Eddington luminosity, making this one of the lowest luminosity soft states recorded in X-ray binaries. We also find that the accretion disk extended towards the compact object during its transition from hard to soft, with the inner radius estimated to be $R_{\mathrm{in}}=28.0^{+0.7}_{-0.4} R_g$ or $\sim12R_g$, dependent on the boundary condition chosen, assuming the above distance and mass, a spectral hardening factor $f=1.7$ and a binary inclination $i=55^{\circ}$.
\end{abstract}

\begin{keywords}
accretion, accretion discs -- black hole physics -- X-rays: binaries -- X-rays: individual: Swift J1753.5-0127.
\end{keywords}

\section{Introduction}

\indent Galactic black hole X-ray transients (BHXRTs) are low-mass X-ray binaries (LMXBs) in which a black hole (BH) accretes material from a donor star via an accretion disk. BHXRTs exhibit distinct spectral states over the course of an outburst, defined by the relative strength of the observed soft and hard continuum components of the X-ray spectrum, as well as the X-ray variability \citep{McClintock-2006,vanderKlis-2006,Belloni-2014}. A typical BHXRT commences an outburst in the hard state, characterised by strong variability ($\sim20-50\%$ rms) and a spectrum dominated by a hard power-law ($\Gamma\sim1.5$) -- believed to originate from an optically thin X-ray ``corona" (see e.g. \citealt{Homan-2001,Done-2007}). At higher luminosities, sources transition to the soft state, during which time the power law component steepens and the spectrum becomes dominated by a quasi-blackbody component -- indicative of a brightening accretion disk (e.g. \citealt{Done-2007}). Accretion states are strongly associated with the formation of jets (in the hard state) and winds (in the soft state) in BHXRTs \citep{Fender-2004,Ponti-2012}.\\
\indent Swift J1753.5-0127 was discovered by the \textit{Swift} Burst Alert Telescope (BAT;  \citealt{Barthelmy-2005}) in 2005 \citep{J1753-ATel} as a hard-spectrum ($\gamma$-ray source) transient at a relatively high Galactic latitude (+12$^{\circ}$). The source luminosity peaked within a week, at a flux of $\sim 200$ mCrab, as observed by the \textit{Rossi X-Ray Timing Explorer} (\textit{RXTE}) All Sky Monitor (ASM; 2--12 keV) \citep{CadolleBel2007}. The source was also detected in the UV, with \textit{Swift}'s Ultraviolet/Optical Telescope (UVOT; \citealt{Still-2005}), and in the radio with MERLIN \citep{Fender-ATel-2005}. A Johnson $R\sim15.8$ optical counterpart was identified by \citet{Halpern-2005}, who noted that it had brightened by at least 5 magnitudes (as it is not visible in the DSS), thereby establishing Swift J1753.5-0127 as a LMXB. Subsequent time-resolved photometry of the optical counterpart revealed $R$-band modulations on a period of $3.2$h, which are indicative of the orbital period ($P_{\mathrm{orb}}$) of the system \citep{Zurita-2008}.\\
\indent Almost immediately after its peak the X-ray flux of Swift J1753.5-0127 started declining, but it then remained roughly constant at $\sim 20$ mCrab (2--12 keV) for over 6 months rather than returning to quiescence as might have been expected for a typical BHXRT\citep{Charles_Coe2006}. The source has still not returned to quiescence, $\sim$10 years after its initial discovery, and has instead exhibited significant long-term ($>400$ day) variability over the course of its prolonged `outburst' \citep{Shaw-2013}. Swift J1753.5-0127 has remained as a persistent LMXB in the hard state for the majority of this time, however it has experienced a number of short-term spectral softenings, characterised by an increase in the inner disk temperature and simultaneous steepening of the power law component \citep{Yoshikawa-2015}. Investigation of the source during one such event with \textit{RXTE} revealed that it had transitioned to the hard intermediate state. However, unlike the majority of BHXRTs, Swift J1753.5-0127 did not continue to the soft state and instead returned to the hard state \citep{Soleri-2013}. The durations of these `failed state transitions' have typically been short ($\sim$25 days), but in early 2015, the source appeared to undergo a more prolonged state transition when the \textit{Swift}-BAT flux dropped to its lowest levels since the source's discovery \citep{Onodera-2015}. Subsequent follow-up with the \textit{Swift} X-ray Telescope (XRT; \citealt{Burrows-2005}) revealed a dramatic increase in the soft band (0.6--2 keV) flux whilst the contribution from the 2--10 keV band had dropped to only $\sim$10\% of the total flux \citep{Shaw-2015}, suggesting that the source was entering a soft state for the first time. \\
\indent With a large ($\Delta R\sim5$ mag) optical increase at outburst, we would not expect to detect any spectroscopic signatures of the donor, due to the optical light being dominated by the accretion disc. \citet{Durant-2009} confirmed this with spectroscopic observations revealing a smooth optical continuum and no evidence for features associated with the donor. With no detectable fluorescence emission either, it has not been possible to obtain any direct evidence of the compact object mass. However, \textit{INTEGRAL} observations highlighted the presence of a hard power-law tail up to $\sim600$ keV, very typical of a black hole candidate (BHC) in the hard state \citep{CadolleBel2007}. Also, the power density spectrum from a pointed \textit{RXTE} observation revealed a $0.6$ Hz quasi-periodic oscillation (QPO) with characteristics typical of BHCs \citep{Morgan-2005}. QPOs have also been seen at 0.08 Hz in optical data \citep{Durant-2009} as well as in a number of X-ray observations after the initial outburst had declined \citep{Ramadevi-2007,CadolleBel2007}. \\
\indent Recently, \citet{Neustroev-2014} reported on evidence for a low mass ($<5M_{\odot}$) BH in Swift J1753.5-0127, based on observations of narrow features in the optical spectrum which they associate with the donor, despite such features not being identifiable or visible in previous spectroscopic studies \citep{Durant-2009}. Given the high Galactic latitude of Swift J1753.5-0127, \citet{CadolleBel2007} concluded that its distance is likely 4--8 kpc. However, in recent work fitting the UV spectrum with an accretion disk model and assuming a $5M_{\odot}$ BH, \citet{Froning-2014} obtain a distance of $<2.8$ kpc and $<3.7$ kpc for a binary inclination of $i=55$ and $0$, respectively. \\
\indent In this paper we present a spectral study of Swift J1753.5-0127 during its 2015 prolonged state transition, which shows characteristics previously unseen in this source. We utilise observations from the \textit{Swift}-XRT and near simultaneous data from \textit{XMM-Newton} \citep{Jansen-2001} and the \textit{Nuclear Spectroscopic Telescope Array} (\textit{NuSTAR}; \citealt{Harrison-2013}). We focus in particular on the emergence of a soft component at a low Eddington fraction, which dominates the 0.7--78 keV spectrum.

\section{Observations and Data Reduction}

\subsection{Swift}
We utilised 173 observations of Swift J1753.5-0127 made with the \textit{Swift} Gamma-ray Burst Mission \citep{Gehrels-2004} between 2005 July 02 and 2015 July 30. \textit{Swift} consists of two pointing telescopes -- the XRT and UVOT \citep{Roming-2005} --  as well as the larger field of view (FoV) coded-mask detector, BAT. For the purposes of this work we have used a number of archival XRT observations (Target IDs: 00030090, 00031232 and 00033140), and we have also monitored the 15--50 keV flux with the BAT. \\
\indent We created a long-term \textit{Swift}-XRT light curve using the online light-curve generator provided by the UK \textit{Swift} Science Data Centre (UKSSDC; \citealt{Evans-2007}), which is presented in Fig. \ref{LC}, along with corresponding light curves from the \textit{Swift}-BAT transient monitor. Hardness ratios were obtained by filtering \textit{Swift}-XRT light curves in ``hard" (1.5--10 keV) and ``soft" (0.3--1.5 keV) energy bands which we use to construct a hardness-intensity diagram (HID, presented in Fig. \ref{HID}. All \textit{Swift}-XRT data were grouped into 1d bins and distinct epochs are represented in the light curves by different symbols and their corresponding hardness ratio (HR) presented in the HID with identical notation.\\
\indent We also extracted spectra of Swift J1753.5-0127 from \textit{Swift}-XRT observations during three epochs. \textit{Swift} observed the source in windowed timing (WT) mode on 2015 March 17, near simultaneous to the \textit{XMM-Newton} observation described below, on 2012 April 24 at the peak of a failed state transition \citep{Yoshikawa-2015} and on 2013 June 18 during the hard state. Data were processed using \scriptsize{HEASOFT} \normalsize v6.16 and count rates were extracted with the task \scriptsize{XSELECT} \normalsize using a circular region 20 pixels in radius ($\approx47''$). The background count rate was extracted from an identically sized source-free region. Spectra were grouped to have a minimum of 20 counts per energy bin. The details of all the spectral observations of Swift J1753.5-0127 are presented in Table \ref{Obs}.

\subsection{XMM-Newton}
A Director's Discretionary Time (DDT) observation of Swift J1753.5-0127 with \textit{XMM-Newton} was performed on 2015 March 19 starting at 02:11 UTC (Obs ID: 0770580201) for a total of 42 ks. For this work we are utilising data from the European Photon Imaging Camera (EPIC) pn detector \citep{Struder-2001}, which operated consecutively in timing (32 ks) and burst (10 ks) modes. Data were reduced using \scriptsize{SAS} \normalsize v14.0.0. \\
\indent The \scriptsize{SAS} \normalsize tool epproc was used to extract the event files from the timing mode data, which were then filtered for flaring particle background and to keep events between RAWX columns 25 and 47. A light curve was extracted with the \scriptsize{SAS} \normalsize task evselect and we found an average count rate of $\sim600$ counts s$^{-1}$. We therefore examined the filtered events for evidence of photon pileup with the task epatplot. Significant pileup was detected in the timing mode data. To reduce the effects of photon pileup, we removed the central five RAWX columns from the data and filtered them to only include single and double pattern events (PATTERN $\leq$4). With the central five columns removed, the average count rate decreased to $\sim115$ counts s$^{-1}$. We extracted source and background spectra with evselect. Response matrices and ancillary files were produced with the \scriptsize SAS \normalsize tools rmfgen and arfgen, respectively. Finally the spectra were binned such that each bin had a minimum signal-to-noise ratio (S/N) greater than 20.\\ 
\indent Immediately following the timing mode observations, EPIC-pn also observed Swift J1753.5-0127 for 10 ks in burst mode, a variation of the standard timing mode which offers high time resolution but with a low duty cycle of 3\%. Spectra were extracted in a similar way to the timing mode data, in RAWY[0:140] as suggested by \citet{Kirsch-2006}, and grouped into energy bins with a minimum S/N of 20. There was no pileup present in the burst mode data. The EPIC-pn covers the energy range from 0.2--10 keV, but the latest EPIC Calibration Technical Note (CAL-TN-0018\footnote{http://xmm2.esac.esa.int/docs/documents/CAL-TN-0018.pdf}) recommends restricting the fit to energies greater than 0.5 keV. To be conservative, our EPIC-pn spectral fits were restricted to the range 0.7--10.0 keV. 

\subsection{NuSTAR}
A DDT observation of Swift J1753.5-0127 was also carried out on 2015 Mar 18 from 19:56 UTC until 2015 March 19 12:06 UTC with \textit{NuSTAR} for a total exposure of 31ks. \textit{NuSTAR} consists of two co-aligned grazing incidence telescopes, FPMA and FPMB, operating in the 3--78 keV energy range. The data were processed with \scriptsize HEASOFT \normalsize v6.16 and the task \scriptsize NUPIPELINE\normalsize. A spectrum and light curve was extracted from a circular region of radius 45$''$. Background counts were extracted from a source-free polygonal region on the same CCD. Spectra were grouped to have a minimum of 50 counts per energy bin. The average count rates were found to be 1.46 and 1.39 counts s$^{-1}$ in FPMA and FPMB, respectively. 

\begin{table*}
\caption{Summary of observations of Swift J1753.5-0127 used for spectral analysis in this work}
\centering
	\begin{tabular}{l l c c c c}
	\hline
	Obs. ID & Start Date (UTC) & MJD & Instrument & Observing Mode & Exposure Time (ks)\\
	\hline
	00031232024 & 2012 April 24 & 56041& \textit{Swift}-XRT & WT & 1.0 \\
	00030090059 & 2013 June 18 & 56461& \textit{Swift}-XRT & WT & 2.0 \\
	00030090067 & 2015 March 17 & 57098 & \textit{Swift}-XRT & WT & 1.0 \\
	80001047002 & 2015 March 18 & 57099 & \textit{NuSTAR} & --- & 31 \\
	0770580201 & 2015 March 19 & 57100 & \textit{XMM-Newton} pn & Timing & 32 \\
	0770580201 & 2015 March 19 & 57100 & \textit{XMM-Newton} pn & Burst & 10 \\
	\hline
	\end{tabular}
\label{Obs}
\end{table*}


\section{Results}

\subsection{The Light Curve and Hardness-Intensity Diagram}
The \textit{Swift}-XRT light curve presented in Fig. \ref{LC} shows the initial 2005 fast rise and exponential decay, which corresponds to the hard state outburst of the source as well as observations of the source in the hard state, characterised by the source's low flux (0.3--10 keV count rate$\sim$10--30 counts s$^{-1}$) and hard spectrum (HR$>0.8$). \\
\indent Also visible in the XRT light curve are two large increases in the 0.3--10 keV count rate (at MJD $\sim$55250 and $\sim$56000), reaching $\sim60\%$ that of the original outburst. Due to the limited coverage by \textit{Swift}-XRT, the timescale of the first of these events is uncertain. However, the second event is simultaneous with a hard X-ray dip present in \textit{Swift}-BAT light curves \citep{Shaw-2013}, which lasted for $\sim25$d. This increase in soft X-rays and simultaneous decrease in hard X-rays is characterised, spectrally, by an increase in inner disk temperature and a decrease in Comptonised emission \citep{Yoshikawa-2015}. A study of the first of these spectral softenings revealed that the source had transitioned to the hard intermediate state, in which there was still significant variability (fractional rms$ >13\%$), but a notably softer spectrum (a failed state transition; \citealt{Soleri-2013}). The two events occupy similar regions in the HID (HR $\sim$0.4--0.5) and exhibit similar spectral properties such as inner disk temperature \citep{Soleri-2013,Yoshikawa-2015}, suggesting that they are both associated with the same phenomena. Following \citet{Soleri-2013}, we shall refer to these two events as the `failed state transitions.' \\
\indent At MJD $\sim$57100 in Fig. \ref{LC} we have highlighted the \textit{Swift}-XRT observations of Swift J1753.5-0127 which took place in March 2015, during which time the \textit{Swift}-BAT flux had been steadily decreasing \citep{Onodera-2015}, visible in the lower panel of Fig. \ref{LC}. The light curve shows a notable increase in 0.3--10 keV count rate ($\sim$55 counts s$^{-1}$ compared to a base rate of $\sim10$--$20$ counts s$^{-1}$). However, the flux does not reach the levels of the failed state transitions. The HR also suggests that the source is much softer than during the failed state transitions (HR $\sim$0.2--0.3, see Fig. \ref{HID}). We can also see the source apparently return to the hard state as the transition progresses, seen as a gradual hardening of the source in Fig. \ref{HID}, coincident with a decrease in 0.3--10 keV count rate and a corresponding increase in \textit{Swift}-BAT flux (Fig. \ref{LC}). In order to investigate the low flux of the apparent soft state we must undertake a detailed study of the X-ray spectrum during the transition.


\subsection{Spectral Fits: \textit{Swift}}
\begin{figure}
\centering
\includegraphics[scale=0.5]{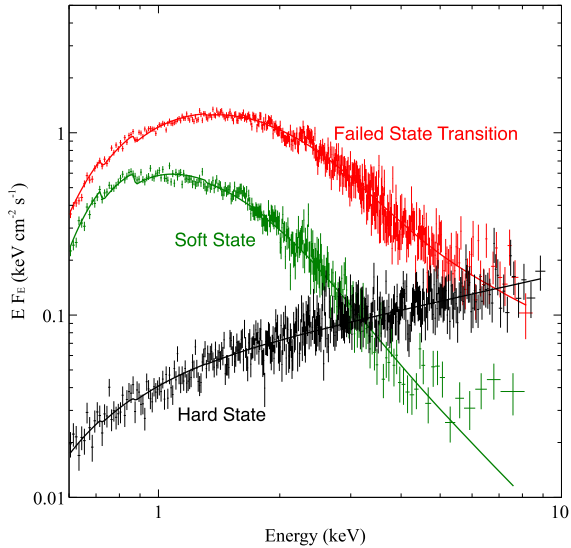}
\caption{A comparison of unfolded \textit{Swift}-XRT spectra from 3 epochs: during the failed state transition (2012 Apr 24; red), during the hard state (2013 June 18; black) and during the apparent new accretion state (2015 March 17; green). The 2012 and 2015 data are fitted with DISKBB+POWERLAW and the 2013 data are fitted with POWERLAW (solid lines).}
\label{Swift}
\end{figure}

All spectral fits were performed using \scriptsize XSPEC \normalsize v12.8.2 \citep{Arnaud-1996} which uses the $\chi^2$ minimisation technique to determine the best fit model. The interstellar absorption is accounted for by the \scriptsize TBABS \normalsize model with \citet{Wilms-2000} abundances and photo-ionisation cross-sections described by \citet{Verner-1996}. We also included a systematic error of $3\%$ for the \textit{Swift}-XRT spectra, given the uncertainties of the response matrix (SWIFT-XRT-CALDB-09\footnote{http://www.swift.ac.uk/analysis/xrt/files/SWIFT-XRT-CALDB-09\_v16.pdf}). We first examined the \textit{Swift}-XRT data in three individual spectral epochs as defined by the HID in Fig. \ref{HID}. As discussed above, we choose one observation each from when the source was in the hard state, at the peak of the failed-state transition and in the apparent new spectral state. The unfolded spectra are presented in Fig. \ref{Swift} and exhibit distinctly different spectral properties. \\
\indent During the hard state, the \textit{Swift}-XRT spectrum can be well described ($\chi^2_{\nu} = 0.96$) by an absorbed power law with photon index $\Gamma=1.51\pm0.04$ (90\% confidence errors are given here and throughout the paper unless otherwise indicated) which is typical of a source in the hard state. During the failed state transition, we find that a one-component model such as an absorbed power law does not fit the data well ($\chi^2_{\nu} = 2.15$). The addition of a multi-colour disk component ($\mathrm{k}T_\mathrm{in}=0.50^{+0.02}_{-0.01}$ keV) to the model provides an acceptable fit ($\chi^2_{\nu} = 0.89$). In this state we find a steeper power law index $\Gamma=3.16^{+0.29}_{-0.32}$. The March 17 \textit{Swift}-XRT spectrum is more complicated to fit. Utilising the \scriptsize DISKBB+POWERLAW \normalsize model, as before, we find a less acceptable fit ($\chi^2_{\nu} = 1.11$; $\mathrm{k}T_\mathrm{in}=0.54^{+0.04}_{-0.05}$ keV and $\Gamma=4.32^{+0.12}_{-0.14}$) although we note that the power law in this spectrum is steeper than during the failed state transition. \\
\indent The March 17 spectrum deviates from the model significantly at energies $>$5 keV, which could be indicative of an additional hard spectral component, though the S/N of the spectrum is too poor to investigate using \textit{Swift}.\\
\indent Figs. \ref{HID} and \ref{Swift} show that the flux of the March 17 \textit{Swift} spectrum appears to be lower than that of the short-lived failed state transition. For the failed state transition the absorbed flux, $F=2.9\times10^{-9}$erg cm$^{-2}$ s$^{-1}$, whilst for the new state $F=1.1\times10^{-9}$erg cm$^{-2}$ s$^{-1}$ (0.6--10 keV) which corresponds to a low Eddington fraction, as discussed further in section \ref{Discussion}. In order to study this apparently low-luminosity soft state, we must investigate the full broadband spectrum.\\

\subsection{Spectral fits: \textit{XMM-Newton} and \textit{NuSTAR}}
\indent For fitting the broadband spectrum, we choose to fix the interstellar column density to $N_H=2.0\times10^{21}$ cm$^{-2}$, obtained by fitting the Ly$\alpha$ absorption line in the UV spectrum \citep{Froning-2014}. We also included a systematic error of 1.2\% and 1\% in the 2--3 and 3--5 keV bands, respectively, to account for uncertainties in the response matrices (see e.g. \citealt{Diaz-Trigo-2014}). Upon investigation of the broadband spectrum, we find that the burst mode spectrum is consistent with the March 17 \textit{Swift}-XRT spectrum below 2 keV, though the timing mode data does not agree with the burst mode data. We therefore utilise the burst mode spectra across the entire energy range (0.7--10 keV), and the timing mode data only at energies $>2$ keV. We also include the \textit{NuSTAR} observations. The cross-normalisation between \textit{XMM-Newton} and \textit{NuSTAR} was allowed to vary with the fit, however it remained close to unity independent of the chosen model. Table \ref{fits} shows the results of the fits of a number of models to the data. The residuals of the model fits are presented in Fig. \ref{residuals}.\\

\subsubsection{Disk blackbody plus powerlaw}
We first consider the \scriptsize DISKBB+POWERLAW \normalsize model, in order to draw comparisons with previous studies of the source in different accretion states \citep{Wilkinson-2009, Yoshikawa-2015, Tomsick-2015}. Such previous studies have shown that the source in both the hard state and the failed state transition is well constrained by a such a model, and we have also noted this above with the \textit{Swift} spectral fits. As suggested by fitting the same model to the \textit{Swift} spectrum (see above), such a simple model cannot accurately represent the 0.7--78 keV spectrum ($\chi^2_{\nu} = 3.02$; see Table \ref{fits} and Fig. \ref{residuals}a). Replacing the power-law with an empirical Comptonisation model (e.g. \scriptsize NTHCOMP\normalsize; \citealt{Zycki-1999}) did not improve the fit ($\chi^2_{\nu} = 3.19$). We note that there is evidence for a slight deviation from the power-law component above $\sim20$ keV, which could be evidence of a weak reflection component, despite the fact that the residuals do not show evidence for a strong Fe emission line at 6.4 keV as might be expected if there was reflection. Adding a reflection component to the power-law with the \scriptsize REFLIONX \normalsize model \citep{Ross-2005} does improve the fit ($\chi^2_{\nu} = 2.08$; Table \ref{fits}), and appears to correct the residuals above 20 keV (Fig. \ref{residuals}b), however, the fit is still not acceptable.\\

\subsubsection{Compton-upscattered disk blackbody}
Due to the increase in the soft component of the spectrum, we utilised the SIMple Power Law model (\scriptsize SIMPL\normalsize; \citealt{Steiner-2009}), a characterisation of Comptonization in which seed photons from an input thermal spectrum are scattered into a power law component. \scriptsize SIMPL \normalsize is designed to be used with soft thermal spectra, which are Compton thin and exhibit a power law index $\Gamma>1$. We utilise the model in the form \scriptsize SIMPL*DISKBB \normalsize. As in the case of the \scriptsize DISKBB+POWERLAW \normalsize model, a two component model does not describe the data well ($\chi^2_{\nu} = 3.08$).\\
\indent As seen from the results in Table \ref{fits} and Fig. \ref{residuals}c the fit improves significantly ($\chi^2_{\nu} = 1.28$), however, if we introduce an additional power law to the model (\scriptsize (SIMPL*DISKBB)+POWERLAW\normalsize). The broad-band unfolded spectrum fitted with this model is presented in Fig. \ref{XMM+Nu}. It is important to note that the disk photons appear to be scattered into a steep power law ($\Gamma_{\mathrm{simpl}}=6.39^{+0.08}_{-0.02}$), whilst there is an additional hard power-law tail ($\Gamma_{\mathrm{PL}}=1.79\pm0.02$). Whilst this models the spectrum more accurately than just a multi-colour disk plus a power law, it is not a self-consistent physical model as it is not clear how to produce such a steep power-law. In order to discuss the underlying physics of the soft state of Swift J1753.5-0127, we must model the additional soft component parametrised here by the additional soft, steep power-law.\\
\indent \scriptsize DISKBB \normalsize is a simple model which equates the disk as a sum of multiple blackbody spectra \citep{Mitsuda-1984} and does not take into account radial and vertical disk structure or relativistic effects. It is possible that these effects may help to explain the excess in the soft component at higher energies which we currently model with a steep power law. Relativistic effects can be considered with the \scriptsize KERRBB \normalsize model, which incorporates a general relativistic disk around a Kerr BH \citep{Li-2005}. We find that replacing \scriptsize DISKBB \normalsize with the \scriptsize KERRBB \normalsize model significantly improves the fit ($\chi^2_{\nu} = 2.24$) from the two component case (\scriptsize DISKBB+POWERLAW\normalsize), however it is still not a formally acceptable fit. We find similar results when introducing the \scriptsize BHSPEC \normalsize model \citep{Davis-2006}, which self-consistently calculates the vertical structure of the disk using stellar atmosphere-like calculations of disk annuli. Utilising the model in the form \scriptsize BHSPEC+POWERLAW \normalsize does not describe the broadband spectrum well ($\chi^2_{\nu} = 2.96$).\\
\indent In either of these more physical models, we cannot accurately reproduce the observed additional soft component in the spectrum. Therefore, we continue to describe the spectral shape using the simpler \scriptsize DISKBB \normalsize model, upscattered into a steep power law component (\scriptsize SIMPL\normalsize).

\subsubsection{Irradiated inner disk}

During the decline of the original outburst of Swift J1753.5-0127, \citet{Chiang-2010} found that the soft region of the X-ray spectrum could not be entirely described by a multi-colour disk due to an additional soft X-ray component emerging. In the new state, we can illustrate this additional soft component with a phenomenological \scriptsize DISKBB+BBODY+POWERLAW \normalsize model. In this scenario, we obtain an improved fit ($\chi^2_{\nu}=1.33$) over two component models, which is characterised by a multi-temperature disk ($\mathrm{k}T_{\mathrm{in}} = 0.29\pm0.01$ keV) with an additional single temperature blackbody ($\mathrm{k}T=0.43\pm0.01$ keV) and a hard power law tail ($\Gamma=1.90\pm0.02$).\\
\indent Following \citet{Chiang-2010}, in which the additional soft X-ray component that emerged was well constrained by irradiation of the inner disk by the Compton tail \citep{Gierlinski-2008}, we apply the irradiated disk model (\scriptsize DISKIR\normalsize) to the 0.7--78 keV spectrum of Swift J1753.5-0127 in its new state. In addition to the irradiated inner disk, \scriptsize DISKIR \normalsize also models the illuminated outer disk, which dominates the optical and UV bands \citep{Gierlinski-2009}. However, as we have not included optical data here, we have fixed the fraction of the flux thermalised in the outer disk, $f_{out}$, to zero. As our spectrum only extends to 78 keV, finally we fix the high energy cutoff (the electron temperature) $kT_e=1000$ keV, due the fit being insensitive to this parameter as well. \\
\indent The resulting fitted parameters are displayed in Table \ref{fits} and show that the fit is significantly improved ($\chi^2_{\nu}=1.34$) compared to two component models, and from Fig. \ref{residuals} we can see that the structure previously visible in simpler fits has been reduced. This is due to the model taking into account illumination of the disk by the Compton tail, which is reprocessed and re-emitted as a quasi-thermal addition to the observed disk emission \citep{Gierlinski-2008}. In the case of Swift J1753.5-0127 we find that the fraction of the Compton tail which is thermalised in this manner is $f_{in}=0.72^{+0.03}_{-0.02}$ and that the ratio of luminosity in the Compton tail to that of the unilluminated disk is $\frac{L_c}{L_d}=7.45^{+0.11}_{-0.10}\times10^{-2}$. 
We also find that the radius of the Compton illuminated region of the disk is $R_{\mathrm{irr}}=1.015\pm0.003R_{\mathrm{in}}$, suggesting that according to the model only the innermost regions of the disk are being irradiated. \\
\indent However, it is recommended \citep{Gierlinski-2008,Gierlinski-2009} to freeze $f_{in}$ to a low value.  This is because the value of $f_{in}$ also relies heavily on assumptions about the reflection albedo of the disk ($f_{in}\sim0.1$ for typical disk albedo of 0.3; \citealt{Gierlinski-2009}). For $f_{in}=0.74$, even if the albedo is only 0.1--0.2 we would infer that $\sim80-90$\% of the sky as seen from the corona is covered by the disk. With such a heavily illuminated disk, we would expect to see strong reflection features in the spectrum. However, despite finding an improved fit if we add a reflection component (\scriptsize REFLIONX\normalsize; $\chi^2_{\nu}=1.23$), there is no strong evidence for reflection in the form of an iron line, nor a Compton hump, as discussed previously. The observed spectrum also shows a weak power law relative to the soft disk emission ($\frac{L_c}{L_d}\sim7\%$), so it seems unlikely that this would provide enough irradiating power to produce such a strong illuminated component. We therefore consider the \scriptsize DISKIR \normalsize model to be an unphysical representation of the spectrum.\\

\subsubsection{Investigation of reflection}
\begin{figure}
\centering
\includegraphics[scale=0.5]{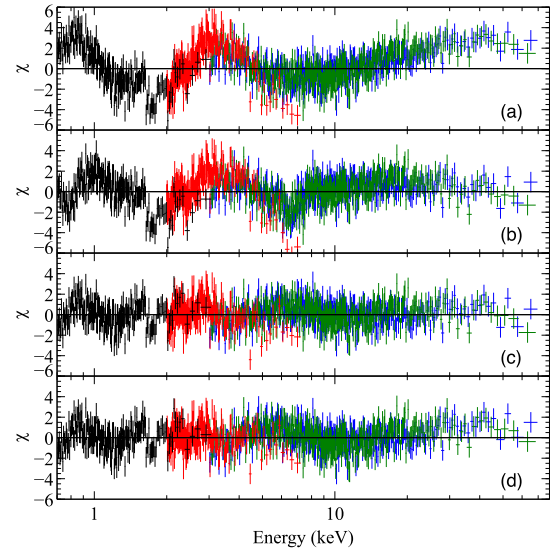}
\caption{$\Delta\chi$ residuals of the best fit model to the combined \textit{XMM-Newton} and \textit{NuSTAR} spectrum of Swift J1753.5-0127. The models are: (a) DISKBB+POWERLAW, (b) DISKBB+POWERLAW+REFLIONX, (c) (SIMPL*DISKBB)+POWERLAW and (d) DISKIR. The figure utilised data from \textit{XMM-Newton}/PN in Burst Mode (black) and Timing Mode (red), \textit{NuSTAR}/FPMA (green) and \textit{NuSTAR}/FPMB (blue).}
\label{residuals}
\end{figure}

\begin{figure}
\centering
\includegraphics[scale=0.5]{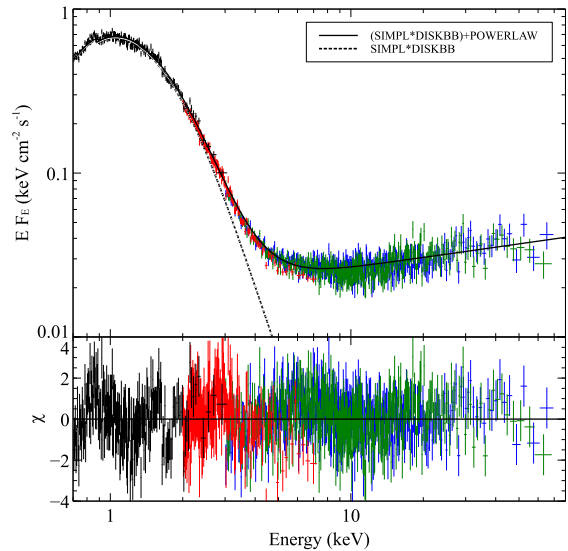}
\caption{The combined \textit{XMM-Newton} and \textit{NuSTAR} spectrum of Swift J1753.5-0127. The 0.7--78 keV spectrum is unfolded and has been fitted with a Compton up-scattered disk model with an additional power law, illustrated by the solid black line. The dotted line represents the SIMPL*DISKBB model fit, without the additional power law. The colour scheme is the same as in Fig. \ref{residuals}.}
\label{XMM+Nu}
\end{figure}

Reflection of hard X-ray photons by the accretion disk is often seen as a combination of fluorescent iron line at 6.4 keV \citep{Fabian-1989} and a broad `Compton hump' which peaks at $\sim30$ keV \citep{Lightman-1988,George-1991}. These features are seen in a number of LMXBs in both the hard and soft states (see e.g. \citealt{Tomsick-2014_2,Furst-2015}). However, in Swift J1753.5-0127 there has been no strong evidence of a reflection component in the X-ray spectrum \citep{Tomsick-2015}, though a broad Fe line has been detected in hard state observations of the source \citep{Hiemstra-2009}. \\
\indent We have already discussed the \scriptsize REFLIONX \normalsize model, which self-consistently includes a narrow Fe line. However in order to examine the presence of an Fe line in detail, we separately include a gaussian at 6.4 keV in the Compton-upscattered disk blackbody model. We find the 90\% upper limit on the equivalent width of a narrow ($\sigma=0.1$ keV) line at 6.4 keV to be $<40$ eV. However, the fit improves ($\Delta\chi^2=54$ for 3 fewer dof) if a broad line is included. The line has a central energy of $E = 6.39^{+0.36}_{-0.23}$ keV, equivalent width $EW = 0.40^{+0.25}_{-0.17}$ keV and an extremely broad width ($\sigma=1.49^{+0.25}_{-0.30}$ keV). The high apparent EW should be associated with a significant reflection Compton hump, however, from the residuals of the \scriptsize(SIMPL*DISKBB)+POWERLAW \normalsize model in Figs. \ref{residuals} and \ref{XMM+Nu}, this feature is not evident. It is possible that the addition of a gaussian is attempting to constrain the unusual continuum shape and upturn in the spectrum at $\sim4-5$ keV with an extremely broad line at $\sim6.4$ keV. \\
\indent However, a broad Fe line may be the result of relativistic broadening. To test this, we use a relativistic convolution of reflection in our model by replacing the hard power law component with the \scriptsize RELXILL \normalsize model \citep{Garcia-2014,Dauser-2014}, which applies relativistic smearing to the complete spectrum. Implementing \scriptsize RELXILL \normalsize provides a formally acceptable fit ($\chi^2_{\nu}=1.24$), but closer inspection of the parameters reveals that the model is perhaps over-fitting the continuum in the region of the Fe line, where there is an upturn in the spectrum (see e.g. the \scriptsize (SIMPL*DISKBB)+POWERLAW \normalsize fit in Fig. \ref{XMM+Nu}). We find that the quality of the \scriptsize RELXILL \normalsize fit is insensitive to a number of important parameters, most notably the BH spin parameter. Another key parameter in the fit is the Fe abundance, which prefers a maximum value of $\mathrm{Fe/solar}=10$. Forcing $\mathrm{Fe/solar}\leq1$ does not improve the fit (compared to $\mathrm{Fe/solar}=10$). It is possible that \scriptsize RELXILL \normalsize is also attempting to constrain the unusual continuum shape as discussed previously in the context of a broad gaussian, hence the unphysical Fe abundance in this scenario. We also find no significant evidence for a Compton hump when applying \scriptsize RELXILL \normalsize to the spectrum and find a reflection fraction of $0.15^{+0.23}_{-0.07}$.

\begin{figure*}
\centering
\includegraphics[scale=0.8, trim = 5mm 0mm 0mm 0mm, clip]{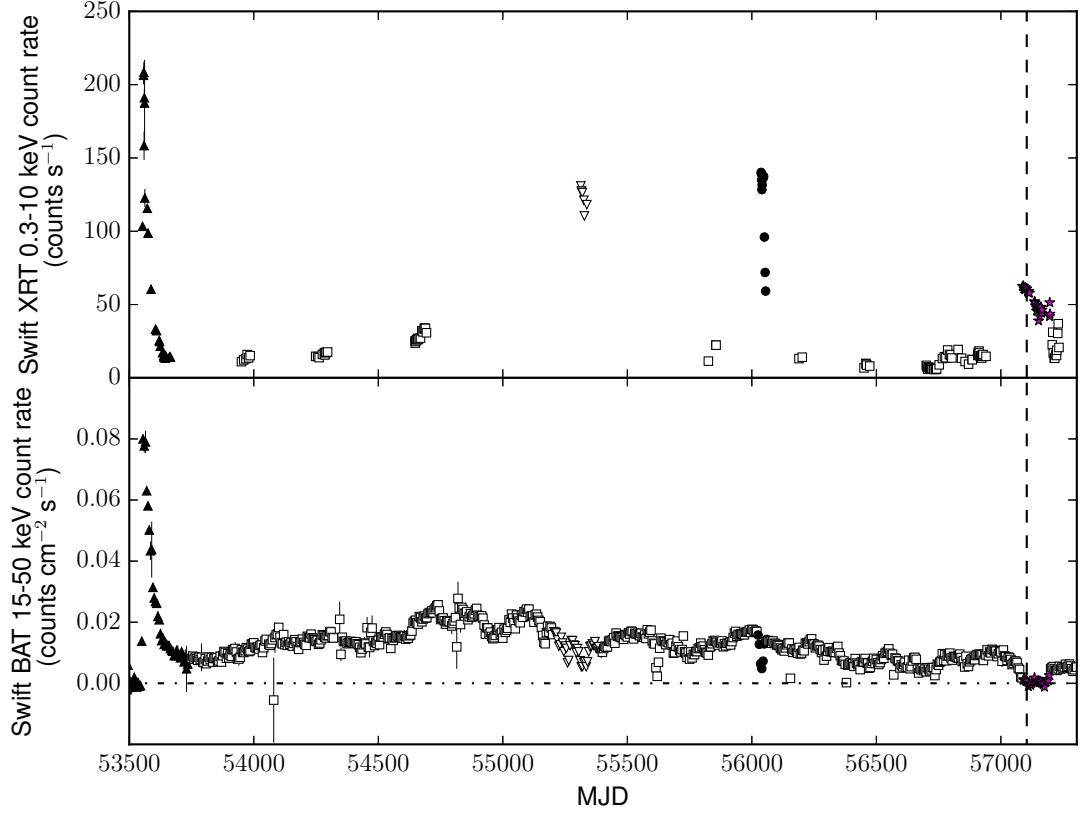}
\caption{\textit{Top}: Long-term light curve of Swift J1753.5-0127 using all available observations with \textit{Swift}-XRT, grouped into 1d bins. \textit{Bottom}: Long-term \textit{Swift}-BAT light curve of Swift J1753.5-0127. Data are grouped into 5d bins. Black triangles represent the initial 2005 outburst and decline, open squares represent observations in the hard state, black circles and open triangles represent observations of the source during two of the failed state transitions, all as observed by \textit{Swift}-XRT. The vertical dashed line represents \textit{Swift}-XRT observations of the source performed between 2015 March 5 and 2015 April 2, during the time of the state transition which is the subject of this work. The dash-dotted line in the bottom panel denotes a \textit{Swift}-BAT count rate of zero.}
\label{LC}
\end{figure*}

\begin{figure}
\centering
\includegraphics[scale=0.45]{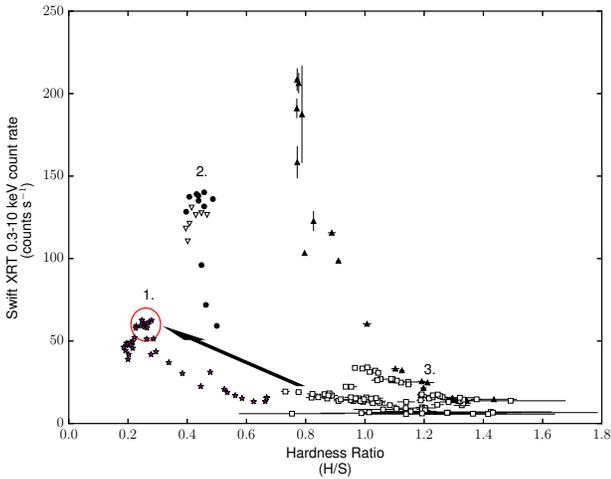}
\caption{Hardness intensity diagram (HID) of Swift J1753.5-0127 using observations from \textit{Swift}-XRT. The data are grouped into 1d bins and the symbols are the same as in Fig. \ref{LC}. The arrow represents the direction of the state transition and the observations performed during the transition are highlighted by the red circle. The hard (H) and soft (S) bands are 1.5--10 keV and 0.3--1.5 keV, respectively. We mark the epochs used to extract the spectra presented in Fig. \ref{Swift} as 1. (soft state), 2. (failed state transition) and 3. (hard state).}
\label{HID}
\end{figure}

\begin{table*}
\caption{Spectral parameters for the combined \textit{XMM-Newton}/\textit{NuSTAR} fits to Swift J1753.5-0127}
\centering
	\begin{tabular}{c c c c c c c}
	\hline
	Parameter & Units & \scriptsize DISKBB+POWERLAW$^a$ \normalsize & \scriptsize DISKBB+POWERLAW+REFLIONX$^{a,b}$ \normalsize & \scriptsize (SIMPL*DISKBB) + POWERLAW \normalsize & \scriptsize DISKIR \normalsize & \scriptsize DISKIR + REFLIONX$^b$ \normalsize \\
	\hline
	$N_H$& $10^{21}$cm$^{-2}$ &  $2.0^c$ & $2.0^c$ & $2.0^c$ & $2.0^c$ & $2.0^c$ \\
	$\mathrm{k}T_\mathrm{in}$ & keV & $\sim0.346$ & $\sim0.363$ & $0.252\pm0.003$ & $0.297\pm0.001$ & $0.302^{+0.002}_{-0.001}$\\
	N$_{\mathrm{disk}}$ & 10$^3$ & $\sim9.55$& $\sim6.85$& $32.8^{+1.6}_{-1.1}$ & $20.0^{+0.1}_{-0.2}$ & $17.9^{+0.2}_{-0.3}$\\
	$\Gamma_{\mathrm{PL}}$ & --- & $\sim2.16$ & --- &1.79$\pm$0.02 & $1.89\pm0.01$ & ---\\
	N$_{\mathrm{PL}}$ & --- & $\sim0.042$ & $\sim0.044$ & $0.017\pm0.001$ & --- & ---\\
	Fe/solar & --- & --- & $\sim20.0$ & --- & --- & $12.1^{+2.5}_{-1.5}$\\
	$\Gamma_{\mathrm{ref}}$ & --- & --- & $\sim2.23$ & --- & --- & $1.98\pm0.01$\\
	$\xi$ & erg cm s$^{-1}$ & --- & $\sim199$& --- & --- & $200^{+1}_{-6}$\\
	N$_{\mathrm{ref}}$ & 10$^{-6}$ & --- & $\sim9.60$ & --- & --- & $2.19^{+0.25}_{-0.18}$\\
	$\Gamma_{\mathrm{simpl}}$ & --- & --- & --- & $6.39^{+0.08}_{-0.02}$ & --- & ---\\
	$f_{SC}$ & --- & --- & --- & $1.00^d$ & --- & ---\\
	$\frac{L_c}{Ld}$ & $10^{-2}$ & --- & --- & --- & $7.45^{+0.11}_{-0.10}$ & $6.57^{+0.14}_{-0.13}$\\
	$f_{in}$ & --- & --- & --- & --- & $0.72^{+0.03}_{-0.02}$ & $0.79^{+0.01}_{-0.03}$\\
	$R_{\mathrm{irr}}$ & $R_{\mathrm{in}}$ & --- & --- & --- & $1.015\pm0.003$ & $1.016\pm0.001$\\
	\hline
	$\chi^2_{\nu}$ & --- & 3132/1037 & 2149/1034 & 1323/1035 & 1391/1035 & 1266/1032\\	
	\hline
	\end{tabular}
\small \\ 
$^a$Due to a poor fit ($\chi^2_{\nu}>2$), accurate (90\% confidence) uncertainties cannot be calculated. \\
$^b$A relativistic interpretation of reflection is discussed in Section 3.3.4.\\
$^c$Fixed. \\
$^d$Fraction of disc photons up scattered reached the upper limit allowable in XSPEC.
\label{fits}
\end{table*}


\section{Discussion}
\label{Discussion}
\subsection{An unusual soft state}
We have presented quasi-simultaneous \textit{XMM-Newton} and \textit{NuSTAR} spectra from the BHC Swift J1753.5-0127 in an accretion state it has never previously been observed in. Best-fit spectral models prefer a multi-colour disk component with a hard Compton tail, with an additional steep ($\Gamma=6.38^{+0.08}_{-0.02}$) power law component which we attempt to reconcile with an irradiated disk model but find that the parameters are unphysical. There is no strong evidence for an Fe emission line in the spectrum at 6.4 keV, nor a reflective Compton hump (if the two power law model is used). Simple, two component models (e.g. \scriptsize DISKBB+POWERLAW\normalsize) did not provide an acceptable fit to the 0.7--78 keV spectrum, and it is evident that in this particular low, soft accretion state, Swift J1753.5-0127 does not exhibit a classical soft state spectrum. Alhough it is not clear how such a steep power law component is produced, physically, this phenomenon is not unique to Swift J1753.5-0127. The BHXRT GRO J1655-40 exhibited an unusual `hypersoft' state during its 2005 outburst in which the photon index was measured to be $\Gamma\simeq6$ \citep{Uttley-2015}. Though the inner disk temperature is much higher ($\mathrm{k}T_\mathrm{in}\simeq1.2$ keV) and the fraction of up-scattered photons much lower ($f_{SC}\simeq0.2$) for the hypersoft state of GRO J1655-40 than the soft state of Swift J1753.5-0127, it nevertheless proves that a steep power law is possible in such sources. \\
\indent The spectrum does, however, show many of the classic characteristics of a LMXB in the soft state -- most significantly the emergence of a disk component in the X-ray spectrum and drop in hard X-ray flux. Separately, we also find that the rms variability has decreased significantly (Uttley et al., in prep) and a sudden drop in the radio flux (Rushton, private communication), suggesting that the compact jet has switched off - also evidence of a transition to a disk-dominated soft state. The HID constructed from $\sim$10 years of \textit{Swift}-XRT observations shows that the source has indeed reached its softest state ever observed, with the HR dropping to $\sim$0.2--0.3, compared with $\sim$0.4 during the failed state transitions studied by \citet{Soleri-2013} and \citet{Yoshikawa-2015}. \\
\indent To investigate the broadband luminosity of the soft state of Swift J1753.5-0127, we estimate the unabsorbed flux (0.1--100 keV) to be 3.5$\times$10$^{-9}$ erg cm$^{-2}$ s$^{-1}$. It must be noted that this calculation is highly sensitive to $N_H$ and is thus only an estimate, due to the uncertainty of the spectral shape outside the 0.7--78 keV range of the observed spectrum. Nevertheless, the chosen energy range is most representative of that required to calculate the intrinsic luminosity of the source. The distance, $d$ to the source is not known and there are a wide range of estimates \citep{CadolleBel2007,Zurita-2008,Froning-2014}. We adopt here the upper limits of $d$=2.8--3.7 kpc (assuming a 5$M_{\odot}$ BH and binary inclination $i<55^{\circ}$) calculated by \citet{Froning-2014} from fitting accretion disk models to the UV spectrum of the source. Assuming a fiducial value of $d$=3 kpc we therefore estimate the luminosity to be 3.8$\times$10$^{36}$erg s$^{-1}$, which, for a 5$M_{\odot}$ BH is 0.60\% of the Eddington limit. For larger BH masses, this value decreases. Soft-to-hard state transitions have been seen to occur in LMXBs at luminosities between 0.3\% and 3\% Eddington \citep{Maccarone-2003b,Kalemci-2013}, though very rarely at the lower end of this range. Swift J1753.5-0127 has therefore likely been observed to be in one of the lowest luminosity soft states recorded in BH systems, third only to those observed in 4U 1630-47 \citep{Tomsick-2014} and the microquasar XTE J1720-318 \citep{Kalemci-2013}. \\
\indent However, to make comparisons with the aforementioned sources, we must estimate the luminosity from the same energy range. For 4U 1630-47, an Eddington fraction was calculated from the 2--10 keV luminosity to be $L/L_{\mathrm{Edd}}=0.008 M^{-1}_{10}\%$ where $M_{10}$ is the mass of the BH in units of 10 $M_{\odot}$ \citep{Tomsick-2014}. We estimate the unabsorbed luminosity of Swift J1753.5-0127 (2--10 keV) to be $L=2.0\times10^{35}$ erg s$^{-1}$ or $L/L_{\mathrm{Edd}}=0.03\%$. This is an order of magnitude higher than 4U 1630-47, though it is still a very low transition luminosity. In the case of XTE J1720-318, \citet{Kalemci-2013} utilised the 3--200 keV flux to estimate an Eddington fraction of $L/L_{\mathrm{Edd}}=0.30\%$ during the soft state. We estimate $L/L_{\mathrm{Edd}}=0.04\%$ for Swift J1753.5-0127 in this range. This is an order of magnitude lower than XTE J1720-318.\\
\indent So far we have seen several cases of unusual soft spectral states at very low luminosities in different sources, of which Swift J1753.5-0127 is now the best-studied example.  Although all these states show soft spectra, they are not like standard disk blackbodies, instead showing very steep but power-law-like tails.  This suggests the possibility that at such low luminosities the disk can persist but its spectrum is modified by strong Compton scattering by plasma with a temperature not much higher than the disk itself, i.e. there is a significant disk `atmosphere.'  This in turn may suggest that disks at such low accretion rates are at the borderline of evaporating, perhaps to form the standard hard state corona (e.g. via the mechanisms described by \citealt{Liu-2006,Meyer-2007,Meyer-Hofmeister-2012}), yet somehow they are able to persist at low luminosities relatively stably in this borderline disk/corona state, compared to disks following the usual soft-hard transition.

\subsection{The inner disk radius}
\indent We can also use the fitting parameters from the \scriptsize (SIMPL*DISKBB)+POWERLAW \normalsize model to estimate the inner radius $R_{\mathrm{in}}$ of the accretion disk of Swift J1753.5-0127. The model normalisation $N_{disk}$ is related to the inner radius as follows:

\begin{equation}
R_{\mathrm{in}}/R_g = \left(0.676 d_{10}f^2 \sqrt{N_{\mathrm{disk}}} \right)/\left(\left( M_{BH}/M_{\odot}\right)\sqrt{\cos i}\right)
\end{equation}

\noindent where $f$ is the spectral hardening factor (see \citealt{Shimura-1995} and Equation 1 of \citealt{Tomsick-2015}) and $d_{10}$ is the distance to the source in units of 10 kpc. For a distance of 3 kpc, $i=55^{\circ}$ (calculated from X-ray spectral analysis of the source in the hard state; \citealt{Reis-2009}), $M_{BH}/M_{\odot}=5$ and $f=1.7$ (a typical value; \citealt{Shimura-1995}) we find $R_{\mathrm{in}}=28.0^{+0.7}_{-0.4}R_g$, though this decreases with increasing $M_{BH}$ and increases for larger distances. Previous studies of the source in the hard state find a heavily truncated disk with $R_{\mathrm{in}}>212R_g$ \citep{Tomsick-2015}. It is important to note that \citet{Tomsick-2015} used a lower inclination ($i=40^{\circ}$), utilising this value in our calculation of $R_{\mathrm{in}}$ would only decrease the value by $\sim4 R_g$. \\
\indent We can separately estimate $R_{\mathrm{in}}$ by considering the friction free boundary condition \citep{Kubota-1998}. From this condition, the inner radius can be calculated as follows:

\begin{equation}
R_{\mathrm{in}}= \sqrt{\frac{3}{7}}\left(\frac{6}{7}\right)^3f^{2}\frac{\sqrt{N_{\mathrm{disk}}}d_{10}}{\sqrt{\cos i}}
\end{equation}

\noindent Using the same values as above, we calculate $R_{\mathrm{in}}=85^{+2}_{-1}$km, which for a $5M_{\odot}$ BH is $\sim12R_{g}$. \\
\indent Regardless of the boundary condition we choose, in the soft state of Swift J1753.5-0127 the inner edge of the disk appears to have moved closer to the BH. In the case of this source, it has not entered a typical soft state, in which the mass accretion rate will reach large Eddington values ($\gtrsim0.1$), consistent with the disk having not yet reached the innermost stable circular orbit (ISCO). However, $R_{\mathrm{in}}$ is significantly closer to the BH than that in the hard state. \\
\indent It is interesting to compare our value of $R_{\mathrm{in}}$ with that of \citet{Yoshikawa-2015}. We find $N_{\mathrm{disk}}$ to be an order of magnitude larger in the soft state than \citet{Yoshikawa-2015} during one of the failed state transitions. This implies that the disk was closer to the BH ($R_{\mathrm{in}}\sim4R_g$ for identical parameters as above) during the short-lived intermediate state, which correlates with the higher disk temperature seen at this time.

\section{Conclusions}
The LMXB Swift J1753.5-0127 transitioned to a low luminosity soft state in March 2015, previously unseen in this source, after $\sim$10 yrs in the hard state. The X-ray spectrum is dominated by a soft accretion disk component and a hard power-law tail, though we find that a simple two component model does not constrain the data well and find that a third component is required. Our best fit model describes the spectrum with a multi-temperature disk with $\mathrm{k}T_\mathrm{in}=0.252$ keV which is scattered into a steep power law ($\Gamma=6.39$), plus an additional, hard ($\Gamma=1.79$) power law tail. We discuss the possibility that the softer, steep power law component could be caused by the irradiation of the inner disk by the Compton tail, but we find that the parameters are unphysical and the fit can be equally well replicated by including an additional multi-temperature disk component in place of the irradiated disk. We find that there is no evidence for Fe emission at 6.4 keV (a 90\% confidence upper limit equivalent width of $<40$ eV), nor for a reflection component in our best fit model. \\
\indent Further studies of the multi-wavelength properties of Swift J1753.5-0127 are being undertaken, including the apparent quenching of the radio jet as the source transitioned to the soft state (Rushtonet al., in prep) and a study of the X-ray variability (Uttley et al., in prep). We will continue to study the source's soft state with a number of instruments. As indicated by the long term \textit{Swift}-BAT light curves in Fig. \ref{LC} and the HID in Fig. \ref{HID}, Swift J1753.5-0127 has returned to the hard state. We will continue to study the source during its return to the hard state.


\section*{Acknowledgements}
The authors thank the referee, Aya Kubota, for helpful suggestions and comments which helped to improve the manuscript. We would like to thank Norbert Schartel and the \textit{XMM-Newton} team for scheduling the ToO observations. The authors are also grateful to the \textit{NuSTAR} team, in particular Fiona Harrison for quickly scheduling contemporaneous ToO observations. The authors would like to thank Daniel Stern for useful discussions. This work made use of data supplied by the UK \textit{Swift} Science Data Centre at the University of Leicester. DA acknowledges support from the Royal Society. This work was supported under NASA Contract No. NNG08FD60C, and made use of data from the {\it NuSTAR} mission, a project led by the California Institute of Technology, managed by the Jet Propulsion Laboratory, and funded by the National Aeronautics and Space Administration. We thank the {\it NuSTAR} Operations, Software and Calibration teams for support with the execution and analysis of these observations. This research has made use of the {\it NuSTAR} Data Analysis Software (NuSTARDAS) jointly developed by the ASI Science Data Center (ASDC, Italy) and the California Institute of Technology (USA). Based on observations obtained with XMM-Newton, an ESA science mission with instruments and contributions directly funded by ESA Member States and NASA.

\footnotesize{
\bibliography{J1753-FINAL.bib}

\begin{thebibliography}{71}
\expandafter\ifx\csname natexlab\endcsname\relax\def\natexlab#1{#1}\fi

\bibitem[{{Arnaud}(1996)}]{Arnaud-1996}
{Arnaud} K.~A., 1996, in Astronomical Society of the Pacific Conference Series,
  Vol. 101, Astronomical Data Analysis Software and Systems V, {Jacoby} G.~H.,
  {Barnes} J., eds., p.~17

\bibitem[{{Barthelmy} {et~al}\mbox{.}(2005){Barthelmy}, {Barbier}, {Cummings},
  {Fenimore}, {Gehrels}, {Hullinger}, {Krimm}, {Markwardt}, {Palmer},
  {Parsons}, {Sato}, {Suzuki}, {Takahashi}, {Tashiro}, \&
  {Tueller}}]{Barthelmy-2005}
{Barthelmy} S.~D. {et~al.}, 2005, \ssr, 120, 143

\bibitem[{{Belloni} \& {Stella}(2014)}]{Belloni-2014}
{Belloni} T.~M., {Stella} L., 2014, \ssr, 183, 43

\bibitem[{{Burrows} {et~al}\mbox{.}(2005){Burrows}, {Hill}, {Nousek}, {Kennea},
  {Wells}, {Osborne}, {Abbey}, {Beardmore}, {Mukerjee}, {Short}, {Chincarini},
  {Campana}, {Citterio}, {Moretti}, {Pagani}, {Tagliaferri}, {Giommi},
  {Capalbi}, {Tamburelli}, {Angelini}, {Cusumano}, {Br{\"a}uninger}, {Burkert},
  \& {Hartner}}]{Burrows-2005}
{Burrows} D.~N. {et~al.}, 2005, \ssr, 120, 165

\bibitem[{{Cabot}, {Wang} \& {Yao}(2013){Cabot}, {Wang}, \& {Yao}}]{Cabot-2013}
{Cabot} S.~H.~C., {Wang} Q.~D., {Yao} Y., 2013, \mnras, 431, 511

\bibitem[{{Cadolle Bel} {et~al}\mbox{.}(2007){Cadolle Bel}, Rib{\'o},
  Rodriguez, Chaty, Corbel, Goldwurm, Frontera, Farinelli, D'Avanzo, Tarana,
  Ubertini, Laurent, Goldoni, \& Mirabel}]{CadolleBel2007}
{Cadolle Bel} M. {et~al.}, 2007, \apj, 659, 549

\bibitem[{{Cassatella}, {Uttley} \& {Maccarone}(2012){Cassatella}, {Uttley}, \&
  {Maccarone}}]{Cassatella-2012}
{Cassatella} P., {Uttley} P., {Maccarone} T.~J., 2012, \mnras, 427, 2985

\bibitem[{{Charles} \& {Coe}(2006)}]{Charles_Coe2006}
{Charles} P.~A., {Coe} M.~J., 2006, in Compact stellar X-ray sources, {Lewin}
  W.~H.~G., {van der Klis} M., eds., Cambridge Univ.Press, p. 215

\bibitem[{{Chiang} {et~al}\mbox{.}(2010){Chiang}, {Done}, {Still}, \&
  {Godet}}]{Chiang-2010}
{Chiang} C.~Y., {Done} C., {Still} M., {Godet} O., 2010, \mnras, 403, 1102

\bibitem[{{Dauser} {et~al}\mbox{.}(2014){Dauser}, {Garc{\'{\i}}a}, {Parker},
  {Fabian}, \& {Wilms}}]{Dauser-2014}
{Dauser} T., {Garc{\'{\i}}a} J., {Parker} M.~L., {Fabian} A.~C., {Wilms} J.,
  2014, \mnras, 444, L100

\bibitem[{{Davis} \& {Hubeny}(2006)}]{Davis-2006}
{Davis} S.~W., {Hubeny} I., 2006, \apjs, 164, 530

\bibitem[{{den Herder} {et~al}\mbox{.}(2001){den Herder}, {Brinkman}, {Kahn},
  {Branduardi-Raymont}, {Thomsen}, {Aarts}, {Audard}, {Bixler}, {den Boggende},
  {Cottam}, {Decker}, {Dubbeldam}, {Erd}, {Goulooze}, {G{\"u}del}, {Guttridge},
  {Hailey}, {Janabi}, {Kaastra}, {de Korte}, {van Leeuwen}, {Mauche},
  {McCalden}, {Mewe}, {Naber}, {Paerels}, {Peterson}, {Rasmussen}, {Rees},
  {Sakelliou}, {Sako}, {Spodek}, {Stern}, {Tamura}, {Tandy}, {de Vries},
  {Welch}, \& {Zehnder}}]{den_Herder-2001}
{den Herder} J.~W. {et~al.}, 2001, \aap, 365, L7

\bibitem[{{D{\'{\i}}az Trigo} {et~al}\mbox{.}(2014){D{\'{\i}}az Trigo},
  {Migliari}, {Miller-Jones}, \& {Guainazzi}}]{Diaz-Trigo-2014}
{D{\'{\i}}az Trigo} M., {Migliari} S., {Miller-Jones} J.~C.~A., {Guainazzi} M.,
  2014, \aap, 571, A76

\bibitem[{{Done}, {Gierli{\'n}ski} \& {Kubota}(2007){Done}, {Gierli{\'n}ski},
  \& {Kubota}}]{Done-2007}
{Done} C., {Gierli{\'n}ski} M., {Kubota} A., 2007, \aapr, 15, 1

\bibitem[{{Durant} {et~al}\mbox{.}(2009){Durant}, {Gandhi}, {Shahbaz},
  {Peralta}, \& {Dhillon}}]{Durant-2009}
{Durant} M., {Gandhi} P., {Shahbaz} T., {Peralta} H.~H., {Dhillon} V.~S., 2009,
  \mnras, 392, 309

\bibitem[{{Evans} {et~al}\mbox{.}(2007){Evans}, {Beardmore}, {Page}, {Tyler},
  {Osborne}, {Goad}, {O'Brien}, {Vetere}, {Racusin}, {Morris}, {Burrows},
  {Capalbi}, {Perri}, {Gehrels}, \& {Romano}}]{Evans-2007}
{Evans} P.~A. {et~al.}, 2007, \aap, 469, 379

\bibitem[{{Fabian} {et~al}\mbox{.}(1989){Fabian}, {Rees}, {Stella}, \&
  {White}}]{Fabian-1989}
{Fabian} A.~C., {Rees} M.~J., {Stella} L., {White} N.~E., 1989, \mnras, 238,
  729

\bibitem[{{Fender}, {Garrington} \& {Muxlow}(2005){Fender}, {Garrington}, \&
  {Muxlow}}]{Fender-ATel-2005}
{Fender} R., {Garrington} S., {Muxlow} T., 2005, ATel, 558, 1

\bibitem[{{Fender}, {Belloni} \& {Gallo}(2004){Fender}, {Belloni}, \&
  {Gallo}}]{Fender-2004}
{Fender} R.~P., {Belloni} T.~M., {Gallo} E., 2004, \mnras, 355, 1105

\bibitem[{{Froning} {et~al}\mbox{.}(2014){Froning}, {Maccarone}, {France},
  {Winter}, {Robinson}, {Hynes}, \& {Lewis}}]{Froning-2014}
{Froning} C.~S., {Maccarone} T.~J., {France} K., {Winter} L., {Robinson} E.~L.,
  {Hynes} R.~I., {Lewis} F., 2014, \apj, 780, 48

\bibitem[{{F{\"u}rst} {et~al}\mbox{.}(2015){F{\"u}rst}, {Nowak}, {Tomsick},
  {Miller}, {Corbel}, {Bachetti}, {Boggs}, {Christensen}, {Craig}, {Fabian},
  {Gandhi}, {Grinberg}, {Hailey}, {Harrison}, {Kara}, {Kennea}, {Madsen},
  {Pottschmidt}, {Stern}, {Walton}, {Wilms}, \& {Zhang}}]{Furst-2015}
{F{\"u}rst} F. {et~al.}, 2015, \apj, 808, 122

\bibitem[{{Garc{\'{\i}}a} {et~al}\mbox{.}(2014){Garc{\'{\i}}a}, {Dauser},
  {Lohfink}, {Kallman}, {Steiner}, {McClintock}, {Brenneman}, {Wilms},
  {Eikmann}, {Reynolds}, \& {Tombesi}}]{Garcia-2014}
{Garc{\'{\i}}a} J. {et~al.}, 2014, \apj, 782, 76

\bibitem[{{Gehrels} {et~al}\mbox{.}(2004){Gehrels}, {Chincarini}, {Giommi},
  {Mason}, {Nousek}, {Wells}, {White}, {Barthelmy}, {Burrows}, {Cominsky},
  {Hurley}, {Marshall}, {M{\'e}sz{\'a}ros}, {Roming}, {Angelini}, {Barbier},
  {Belloni}, {Campana}, {Caraveo}, {Chester}, {Citterio}, {Cline}, {Cropper},
  {Cummings}, {Dean}, {Feigelson}, {Fenimore}, {Frail}, {Fruchter}, {Garmire},
  {Gendreau}, {Ghisellini}, {Greiner}, {Hill}, {Hunsberger}, {Krimm},
  {Kulkarni}, {Kumar}, {Lebrun}, {Lloyd-Ronning}, {Markwardt}, {Mattson},
  {Mushotzky}, {Norris}, {Osborne}, {Paczynski}, {Palmer}, {Park}, {Parsons},
  {Paul}, {Rees}, {Reynolds}, {Rhoads}, {Sasseen}, {Schaefer}, {Short},
  {Smale}, {Smith}, {Stella}, {Tagliaferri}, {Takahashi}, {Tashiro},
  {Townsley}, {Tueller}, {Turner}, {Vietri}, {Voges}, {Ward}, {Willingale},
  {Zerbi}, \& {Zhang}}]{Gehrels-2004}
{Gehrels} N. {et~al.}, 2004, \apj, 611, 1005

\bibitem[{{George} \& {Fabian}(1991)}]{George-1991}
{George} I.~M., {Fabian} A.~C., 1991, \mnras, 249, 352

\bibitem[{{Gierli{\'n}ski}, {Done} \& {Page}(2008){Gierli{\'n}ski}, {Done}, \&
  {Page}}]{Gierlinski-2008}
{Gierli{\'n}ski} M., {Done} C., {Page} K., 2008, \mnras, 388, 753

\bibitem[{{Gierli{\'n}ski}, {Done} \& {Page}(2009){Gierli{\'n}ski}, {Done}, \&
  {Page}}]{Gierlinski-2009}
{Gierli{\'n}ski} M., {Done} C., {Page} K., 2009, \mnras, 392, 1106

\bibitem[{{Halpern}(2005)}]{Halpern-2005}
{Halpern} J.~P., 2005, ATel, 549, 1

\bibitem[{{Harrison} {et~al}\mbox{.}(2013){Harrison}, {Craig}, {Christensen},
  {Hailey}, {Zhang}, {Boggs}, {Stern}, {Cook}, {Forster}, {Giommi},
  {Grefenstette}, {Kim}, {Kitaguchi}, {Koglin}, {Madsen}, {Mao}, {Miyasaka},
  {Mori}, {Perri}, {Pivovaroff}, {Puccetti}, {Rana}, {Westergaard}, {Willis},
  {Zoglauer}, {An}, {Bachetti}, {Barri{\`e}re}, {Bellm}, {Bhalerao},
  {Brejnholt}, {Fuerst}, {Liebe}, {Markwardt}, {Nynka}, {Vogel}, {Walton},
  {Wik}, {Alexander}, {Cominsky}, {Hornschemeier}, {Hornstrup}, {Kaspi},
  {Madejski}, {Matt}, {Molendi}, {Smith}, {Tomsick}, {Ajello}, {Ballantyne},
  {Balokovi{\'c}}, {Barret}, {Bauer}, {Blandford}, {Brandt}, {Brenneman},
  {Chiang}, {Chakrabarty}, {Chenevez}, {Comastri}, {Dufour}, {Elvis}, {Fabian},
  {Farrah}, {Fryer}, {Gotthelf}, {Grindlay}, {Helfand}, {Krivonos}, {Meier},
  {Miller}, {Natalucci}, {Ogle}, {Ofek}, {Ptak}, {Reynolds}, {Rigby},
  {Tagliaferri}, {Thorsett}, {Treister}, \& {Urry}}]{Harrison-2013}
{Harrison} F.~A. {et~al.}, 2013, \apj, 770, 103

\bibitem[{{Hiemstra} {et~al}\mbox{.}(2009){Hiemstra}, {Soleri}, {M{\'e}ndez},
  {Belloni}, {Mostafa}, \& {Wijnands}}]{Hiemstra-2009}
{Hiemstra} B., {Soleri} P., {M{\'e}ndez} M., {Belloni} T., {Mostafa} R.,
  {Wijnands} R., 2009, \mnras, 394, 2080

\bibitem[{{Homan} {et~al}\mbox{.}(2001){Homan}, {Wijnands}, {van der Klis},
  {Belloni}, {van Paradijs}, {Klein-Wolt}, {Fender}, \&
  {M{\'e}ndez}}]{Homan-2001}
{Homan} J., {Wijnands} R., {van der Klis} M., {Belloni} T., {van Paradijs} J.,
  {Klein-Wolt} M., {Fender} R., {M{\'e}ndez} M., 2001, \apjs, 132, 377

\bibitem[{{Jansen} {et~al}\mbox{.}(2001){Jansen}, {Lumb}, {Altieri}, {Clavel},
  {Ehle}, {Erd}, {Gabriel}, {Guainazzi}, {Gondoin}, {Much}, {Munoz}, {Santos},
  {Schartel}, {Texier}, \& {Vacanti}}]{Jansen-2001}
{Jansen} F. {et~al.}, 2001, \aap, 365, L1

\bibitem[{{Kalemci} {et~al}\mbox{.}(2013){Kalemci}, {Din{\c c}er}, {Tomsick},
  {Buxton}, {Bailyn}, \& {Chun}}]{Kalemci-2013}
{Kalemci} E., {Din{\c c}er} T., {Tomsick} J.~A., {Buxton} M.~M., {Bailyn}
  C.~D., {Chun} Y.~Y., 2013, \apj, 779, 95

\bibitem[{{Kirsch} {et~al}\mbox{.}(2006){Kirsch}, {Sch{\"o}nherr},
  {Kendziorra}, {Freyberg}, {Martin}, {Wilms}, {Mukerjee}, {Breitfellner},
  {Smith}, \& {Staubert}}]{Kirsch-2006}
{Kirsch} M.~G.~F. {et~al.}, 2006, \aap, 453, 173

\bibitem[{{Kubota} {et~al}\mbox{.}(1998){Kubota}, {Tanaka}, {Makishima},
  {Ueda}, {Dotani}, {Inoue}, \& {Yamaoka}}]{Kubota-1998}
{Kubota} A., {Tanaka} Y., {Makishima} K., {Ueda} Y., {Dotani} T., {Inoue} H.,
  {Yamaoka} K., 1998, \pasj, 50, 667

\bibitem[{{Li} {et~al}\mbox{.}(2005){Li}, {Zimmerman}, {Narayan}, \&
  {McClintock}}]{Li-2005}
{Li} L.-X., {Zimmerman} E.~R., {Narayan} R., {McClintock} J.~E., 2005, \apjs,
  157, 335

\bibitem[{{Lightman} \& {White}(1988)}]{Lightman-1988}
{Lightman} A.~P., {White} T.~R., 1988, \apj, 335, 57

\bibitem[{{Liu}, {Meyer} \& {Meyer-Hofmeister}(2006){Liu}, {Meyer}, \&
  {Meyer-Hofmeister}}]{Liu-2006}
{Liu} B.~F., {Meyer} F., {Meyer-Hofmeister} E., 2006, \aap, 454, L9

\bibitem[{{Maccarone}(2003)}]{Maccarone-2003b}
{Maccarone} T.~J., 2003, \aap, 409, 697

\bibitem[{{McClintock} \& {Remillard}(2006)}]{McClintock-2006}
{McClintock} J.~E., {Remillard} R.~A., 2006, in Compact stellar X-ray sources,
  {Lewin} W.~H.~G., {van der Klis} M., eds., Cambridge Univ.Press, p. 157

\bibitem[{{Meyer}, {Liu} \& {Meyer-Hofmeister}(2007){Meyer}, {Liu}, \&
  {Meyer-Hofmeister}}]{Meyer-2007}
{Meyer} F., {Liu} B.~F., {Meyer-Hofmeister} E., 2007, \aap, 463, 1

\bibitem[{{Meyer-Hofmeister}, {Liu} \& {Meyer}(2012){Meyer-Hofmeister}, {Liu},
  \& {Meyer}}]{Meyer-Hofmeister-2012}
{Meyer-Hofmeister} E., {Liu} B.~F., {Meyer} F., 2012, \aap, 544, A87

\bibitem[{{Miller}, {Homan} \& {Miniutti}(2006){Miller}, {Homan}, \&
  {Miniutti}}]{Miller-2006}
{Miller} J.~M., {Homan} J., {Miniutti} G., 2006, \apjl, 652, L113

\bibitem[{{Mitsuda} {et~al}\mbox{.}(1984){Mitsuda}, {Inoue}, {Koyama},
  {Makishima}, {Matsuoka}, {Ogawara}, {Suzuki}, {Tanaka}, {Shibazaki}, \&
  {Hirano}}]{Mitsuda-1984}
{Mitsuda} K. {et~al.}, 1984, \pasj, 36, 741

\bibitem[{{Morgan} {et~al}\mbox{.}(2005){Morgan}, {Swank}, {Markwardt}, \&
  {Gehrels}}]{Morgan-2005}
{Morgan} E., {Swank} J., {Markwardt} C., {Gehrels} N., 2005, ATel, 550, 1

\bibitem[{{Neustroev} {et~al}\mbox{.}(2014){Neustroev}, {Veledina}, {Poutanen},
  {Zharikov}, {Tsygankov}, {Sjoberg}, \& {Kajava}}]{Neustroev-2014}
{Neustroev} V.~V., {Veledina} A., {Poutanen} J., {Zharikov} S.~V., {Tsygankov}
  S.~S., {Sjoberg} G., {Kajava} J.~J.~E., 2014, \mnras, 445, 2424

\bibitem[{{Onodera} {et~al}\mbox{.}(2015){Onodera}, {Negoro}, {Nakahira},
  {Ueno}, {Tomida}, {Kimura}, {Ishikawa}, {Nakagawa}, {Mihara}, {Sugizaki},
  {Morii}, {Serino}, {Sugimoto}, {Takagi}, {Yoshikawa}, {Matsuoka}, {Kawai},
  {Tachibana}, {Yoshii}, {Yoshida}, {Sakamoto}, {Kawakubo}, {Ohtsuki},
  {Tsunemi}, {Uchida}, {Nakajima}, {Fukushima}, {Suzuki}, {Fujita}, {Honda},
  {Namba}, {Ueda}, {Shidatsu}, {Kawamuro}, {Hori}, {Tsuboi}, {Kawagoe},
  {Yamauchi}, {Morooka}, {Itoh}, \& {Yamaoka}}]{Onodera-2015}
{Onodera} T. {et~al.}, 2015, The Astronomer's Telegram, 7196, 1

\bibitem[{{Palmer} {et~al}\mbox{.}(2005){Palmer}, {Barthelmey}, {Cummings},
  {Gehrels}, {Krimm}, {Markwardt}, {Sakamoto}, \& {Tueller}}]{J1753-ATel}
{Palmer} D.~M., {Barthelmey} S.~D., {Cummings} J.~R., {Gehrels} N., {Krimm}
  H.~A., {Markwardt} C.~B., {Sakamoto} T., {Tueller} J., 2005, ATel, 546, 1

\bibitem[{{Ponti} {et~al}\mbox{.}(2012){Ponti}, {Fender}, {Begelman}, {Dunn},
  {Neilsen}, \& {Coriat}}]{Ponti-2012}
{Ponti} G., {Fender} R.~P., {Begelman} M.~C., {Dunn} R.~J.~H., {Neilsen} J.,
  {Coriat} M., 2012, \mnras, 422, L11

\bibitem[{Ramadevi \& Seetha(2007)}]{Ramadevi-2007}
Ramadevi M.~C., Seetha S., 2007, \mnras, 378, 182

\bibitem[{{Reis} {et~al}\mbox{.}(2009){Reis}, {Fabian}, {Ross}, \&
  {Miller}}]{Reis-2009}
{Reis} R.~C., {Fabian} A.~C., {Ross} R.~R., {Miller} J.~M., 2009, \mnras, 395,
  1257

\bibitem[{{Reynolds} {et~al}\mbox{.}(2010){Reynolds}, {Miller}, {Homan}, \&
  {Miniutti}}]{Reynolds-2010}
{Reynolds} M.~T., {Miller} J.~M., {Homan} J., {Miniutti} G., 2010, \apj, 709,
  358

\bibitem[{{Roming} {et~al}\mbox{.}(2005){Roming}, {Kennedy}, {Mason}, {Nousek},
  {Ahr}, {Bingham}, {Broos}, {Carter}, {Hancock}, {Huckle}, {Hunsberger},
  {Kawakami}, {Killough}, {Koch}, {McLelland}, {Smith}, {Smith}, {Soto},
  {Boyd}, {Breeveld}, {Holland}, {Ivanushkina}, {Pryzby}, {Still}, \&
  {Stock}}]{Roming-2005}
{Roming} P.~W.~A. {et~al.}, 2005, \ssr, 120, 95

\bibitem[{{Ross} \& {Fabian}(2005)}]{Ross-2005}
{Ross} R.~R., {Fabian} A.~C., 2005, \mnras, 358, 211

\bibitem[{{Shaw} {et~al}\mbox{.}(2013){Shaw}, {Charles}, {Bird}, {Cornelisse},
  {Casares}, {Lewis}, {Mu{\~n}oz-Darias}, {Russell}, \& {Zurita}}]{Shaw-2013}
{Shaw} A.~W. {et~al.}, 2013, \mnras, 433, 740

\bibitem[{{Shaw} {et~al}\mbox{.}(2015){Shaw}, {Charles}, {Gandhi}, \&
  {Altamirano}}]{Shaw-2015}
{Shaw} A.~W., {Charles} P.~A., {Gandhi} P., {Altamirano} D., 2015, The
  Astronomer's Telegram, 7216, 1

\bibitem[{{Shimura} \& {Takahara}(1995)}]{Shimura-1995}
{Shimura} T., {Takahara} F., 1995, \apj, 445, 780

\bibitem[{{Soleri} {et~al}\mbox{.}(2013){Soleri}, {Mu{\~n}oz-Darias}, {Motta},
  {Belloni}, {Casella}, {M{\'e}ndez}, {Altamirano}, {Linares}, {Wijnands},
  {Fender}, \& {van der Klis}}]{Soleri-2013}
{Soleri} P. {et~al.}, 2013, \mnras, 429, 1244

\bibitem[{{Steiner} {et~al}\mbox{.}(2009){Steiner}, {Narayan}, {McClintock}, \&
  {Ebisawa}}]{Steiner-2009}
{Steiner} J.~F., {Narayan} R., {McClintock} J.~E., {Ebisawa} K., 2009, \pasp,
  121, 1279

\bibitem[{{Still} {et~al}\mbox{.}(2005){Still}, {Roming}, {Brocksopp}, \&
  {Markwardt}}]{Still-2005}
{Still} M., {Roming} P., {Brocksopp} C., {Markwardt} C.~B., 2005, ATel, 553, 1

\bibitem[{{Str{\"u}der} {et~al}\mbox{.}(2001){Str{\"u}der}, {Briel}, {Dennerl},
  {Hartmann}, {Kendziorra}, {Meidinger}, {Pfeffermann}, {Reppin}, {Aschenbach},
  {Bornemann}, {Br{\"a}uninger}, {Burkert}, {Elender}, {Freyberg}, {Haberl},
  {Hartner}, {Heuschmann}, {Hippmann}, {Kastelic}, {Kemmer}, {Kettenring},
  {Kink}, {Krause}, {M{\"u}ller}, {Oppitz}, {Pietsch}, {Popp}, {Predehl},
  {Read}, {Stephan}, {St{\"o}tter}, {Tr{\"u}mper}, {Holl}, {Kemmer}, {Soltau},
  {St{\"o}tter}, {Weber}, {Weichert}, {von Zanthier}, {Carathanassis}, {Lutz},
  {Richter}, {Solc}, {B{\"o}ttcher}, {Kuster}, {Staubert}, {Abbey}, {Holland},
  {Turner}, {Balasini}, {Bignami}, {La Palombara}, {Villa}, {Buttler},
  {Gianini}, {Lain{\'e}}, {Lumb}, \& {Dhez}}]{Struder-2001}
{Str{\"u}der} L. {et~al.}, 2001, \aap, 365, L18

\bibitem[{{Tomsick} {et~al}\mbox{.}(2014{\natexlab{a}}){Tomsick}, {Nowak},
  {Parker}, {Miller}, {Fabian}, {Harrison}, {Bachetti}, {Barret}, {Boggs},
  {Christensen}, {Craig}, {Forster}, {F{\"u}rst}, {Grefenstette}, {Hailey},
  {King}, {Madsen}, {Natalucci}, {Pottschmidt}, {Ross}, {Stern}, {Walton},
  {Wilms}, \& {Zhang}}]{Tomsick-2014_2}
{Tomsick} J.~A. {et~al.}, 2014{\natexlab{a}}, \apj, 780, 78

\bibitem[{{Tomsick} {et~al}\mbox{.}(2015){Tomsick}, {Rahoui}, {Kolehmainen},
  {Miller-Jones}, {F{\"u}rst}, {Yamaoka}, {Akitaya}, {Corbel}, {Coriat},
  {Done}, {Gandhi}, {Harrison}, {Huang}, {Kaaret}, {Kalemci}, {Kanda},
  {Migliari}, {Miller}, {Moritani}, {Stern}, {Uemura}, \&
  {Urata}}]{Tomsick-2015}
{Tomsick} J.~A. {et~al.}, 2015, \apj, 808, 85

\bibitem[{{Tomsick} {et~al}\mbox{.}(2014{\natexlab{b}}){Tomsick}, {Yamaoka},
  {Corbel}, {Kalemci}, {Migliari}, \& {Kaaret}}]{Tomsick-2014}
{Tomsick} J.~A., {Yamaoka} K., {Corbel} S., {Kalemci} E., {Migliari} S.,
  {Kaaret} P., 2014{\natexlab{b}}, \apj, 791, 70

\bibitem[{{Uttley} \& {Klein-Wolt}(2015)}]{Uttley-2015}
{Uttley} P., {Klein-Wolt} M., 2015, \mnras, 451, 475

\bibitem[{{van der Klis}(2006)}]{vanderKlis-2006}
{van der Klis} M., 2006, in Compact stellar X-ray sources, {Lewin} W.~H.~G.,
  {van der Klis} M., eds., p.~39

\bibitem[{{Verner} {et~al}\mbox{.}(1996){Verner}, {Ferland}, {Korista}, \&
  {Yakovlev}}]{Verner-1996}
{Verner} D.~A., {Ferland} G.~J., {Korista} K.~T., {Yakovlev} D.~G., 1996, \apj,
  465, 487

\bibitem[{{Wilkinson} \& {Uttley}(2009)}]{Wilkinson-2009}
{Wilkinson} T., {Uttley} P., 2009, \mnras, 397, 666

\bibitem[{{Wilms}, {Allen} \& {McCray}(2000){Wilms}, {Allen}, \&
  {McCray}}]{Wilms-2000}
{Wilms} J., {Allen} A., {McCray} R., 2000, \apj, 542, 914

\bibitem[{{Yoshikawa} {et~al}\mbox{.}(2015){Yoshikawa}, {Yamada}, {Nakahira},
  {Matsuoka}, {Negoro}, {Mihara}, \& {Tamagawa}}]{Yoshikawa-2015}
{Yoshikawa} A., {Yamada} S., {Nakahira} S., {Matsuoka} M., {Negoro} H.,
  {Mihara} T., {Tamagawa} T., 2015, \pasj, 67, 11

\bibitem[{{Zurita} {et~al}\mbox{.}(2008){Zurita}, {Durant}, {Torres},
  {Shahbaz}, {Casares}, \& {Steeghs}}]{Zurita-2008}
{Zurita} C., {Durant} M., {Torres} M.~A.~P., {Shahbaz} T., {Casares} J.,
  {Steeghs} D., 2008, \apj, 681, 1458

\bibitem[{{{\.Z}ycki}, {Done} \& {Smith}(1999){{\.Z}ycki}, {Done}, \&
  {Smith}}]{Zycki-1999}
{{\.Z}ycki} P.~T., {Done} C., {Smith} D.~A., 1999, \mnras, 309, 561

\end{thebibliography}
}
\end{document}